\documentclass[aps, 12pt, amsfonts, amssymb, amsmath]{revtex4}

\usepackage{graphicx}

\newcommand{\dpartial}[2]{\frac{\partial #1}{\partial #2}}
\newcommand{\dtotal}[2]{\frac{d #1}{d #2}}

\begin{document}

\title{Dynamics  of shear-transformation zones in amorphous plasticity:
non-linear theory at low temperatures}

\author{Leonid Pechenik} \affiliation{Department of Physics,
University of California, Santa Barbara, CA  93106-9530  USA}
\date{July, 2004}

\begin{abstract}
We use considerations of energy balance and dissipation to derive a
self-consistent  version of the shear-transformation-zone (STZ) theory
of plastic deformation in amorphous solids.  The theory is generalized
to include arbitrary spatial orientations of STZs. Continuum
equations for elasto-plastic material and their energy
balance properties are discussed.
\end{abstract}
\maketitle

\numberwithin{equation}{section}

\section{Introduction}\label{sec:1}

Important progress has been made in a recent series of papers by Falk, 
Langer and myself 
on the shear-transformation-zone (STZ) theory of plastic deformation in 
amorphous solids \cite{LANGER:2003, FALK:2004}.
In the first paper in this series we introduced and explored
an energetic approach to the STZ theory at temperatures far
below glass transition temperature, which helped us to define the 
limits of the theory's form. 
The finite-temperature version of the theory developed in 
the second paper \cite{FALK:2004}
allowed us to make predictions that were comparable to
to experimental observations of the behavior of bulk
metallic glasses (Kato et al.\cite{KATO:1998}, Lu et al.\cite{LU:2003}).
The success and the questions that these studies posed
prompt us to look more carefully at the fundamentals of the theory and
understand the extent to which the simple approximations that we used
were correct, and how to construct the theory without them.
This paper is focused on further generalizing and expanding 
the low-temperature STZ theory of plasticity.
In particular, we reexamine the physical significance of two
parameters that occurred in the energy balance equations 
introduced in Ref. \cite{LANGER:2003}; and we show explicitly how to derive
the tensorial version of the theory, already used in Ref. \cite{EASTGATE:2003}, 
that is needed in order to describe situations in which the orientation
of the stress changes as a function of position and time.
Finally, for completeness, we derive a full set of elasto-plastic continuum
equations of motion for this class of models. 

The STZ theory of plasticity of amorphous materials at low temperatures 
was proposed by Falk and Langer in \cite{FALK:1998}.
It is based on the previous works of Cohen, Turnbull, Spaepen, Argon 
\cite{TURNBULL:1970, SPAEPEN:1977,ARGON:1979}, which argued 
that non-crystalline solids
plasticity is due to atomic rearrangements at localized sites. This
picture has  also been confirmed by a number of computational studies
\cite{SROLOVITZ:1981, DENG:1989}.
However, unlike the earlier theories, the STZ theory focuses in detail on
how rearrangements at the localized sites (shear-transformation-zones)
occur, and identifies as important dynamical variables not only
the concentration of the STZs, but also their orientations.
This new variable allowed  immediately to obtain a 
description of elastic and plastic behavior 
as an exchange of stability between the two steady states.
Such a simple mathematical treatment appears to us to be much more 
natural than the approach of traditional plasticity theory with its 
yield criteria.

Moreover, the original STZ theory offered an explanation of a wide class of  
plasticity phenomena such as work hardening, strain softening,  
the Bauschinger effect, and others. But, as pointed out in \cite{FALK:1998}, 
it had an inconsistency which implied that the proposed form was not completely
correct. The energetic approach introduced in \cite{LANGER:2003} allowed to correct 
the inconsistency for a simple case of quasilinear approximation.
As shown there, even in such a simple form the STZ theory captured the important 
features observed in both mechanical tests and calorimetric 
measurements of glassy polymers at temperatures far below glass transition 
temperature \cite{HASAN:1993}. A generalization of such an approach (also for
the quasilinear approximation) to higher
temperatures \cite{FALK:2004} has proven to be quantitatively successful
in the description of the viscoelastic response of bulk metallic glasses 
under tensile loading\cite{LU:2003, KATO:1998}.
However, as argued in  \cite{LANGER:2003,FALK:2004, FALK:2000}, the application of
quasilinear theory is limited. Most notably, the quasilinear approximation 
exaggerates plastic flow at small stresses and low temperatures, 
and reduces memory effects.
As only the non-linear STZ theory can be expected to adequately describe 
molecular rearrangements, it must be further developed
in order to reach precise quantitative agreement with experiment.
One of the purposes of this paper is to expand the energetic approach
introduced in \cite{LANGER:2003} to the non-linear STZ theory.

A major challenge in developing the STZ theory
was defining the form of the STZ creation and annihilation rates.
In the original paper \cite{FALK:1998} these rates  were proposed to be 
proportional to the rate of 
plastic work $\sigma \dot{\varepsilon}_{pl}$. Since this work 
can become negative, it was obvious that this form was not acceptable
(the inconsistency noted above). An easy (but artificial) remedy was
proposed in \cite{EASTGATE:2003} -- to make them proportional to 
the absolute value $|\sigma \dot{\varepsilon}_{pl}|$. This was sufficient
for handling complicated numerical simulations of necking where 
$\sigma \dot{\varepsilon}_{pl}$ becomes negative during unloading.
These simulations also explicitly demonstrated
that thinning of the neck could continue even after stretching of
the sample had been stopped. This raised a question whether the
proposed forms of the STZ theory agreed with fundamental physical 
principles -- the first and second laws of thermodynamics.
It appeared that, indeed, they did -- the plastic deformation 
of the neck was driven by the energy stored in the bulk of material, 
and this process was dissipative.

Beyond that, the involvement of energy concepts in the consideration of
the theory opened a different perspective. In this paper we make a conjecture
that will be the basis of all of the following discussion -- that 
creation and annihilation rates are proportional to the {\em 
rate of energy dissipation}.
This conjecture of proportionality allows us to self-consistently
define all components of the theory (Section \ref{sec:2}). 
The formalism developed there is a useful tool in limiting the arbitrariness 
of possible forms of the dynamical equations, transcending the current 
framework of low-temperature STZ theory. In Section \ref{sec:3examples} 
we demonstrate conclusions of Section \ref{sec:2} on two important examples. 

Another significant limitation of the original STZ theory was that
it considered STZs oriented in a single direction only. 
In earlier studies that had to deal with stress changing its direction 
\cite{FALK-THESIS, LANGER:2000, LANGER:2001, EASTGATE:2003}, 
the form of the theory for amorphous material, isotropic in its nature, 
had to be guessed  on a phenomenological basis. In Section \ref{sec:3} 
of this paper
we return to the microscopic basics and construct a theory that includes
STZs oriented in all possible directions. Thereafter, 
we introduce an approximation that allows us to 
rewrite the theory in a simpler tensorial form, with order 
parameters being the first and second moments of the orientational 
density of the STZs. This tensorial form is comparable to the 
above mentioned phenomenological theory.

In Section \ref{sec:4} we combine ideas of the previous sections,
applying the energetic approach from Section \ref{sec:2} to the 
isotropic model of STZ theory from Section \ref{sec:3}.

An understanding of energetic processes in the plastic degrees of
freedom allows us to deal more carefully with spatially  distributed
systems, which  we discuss in Section \ref{sec:5}.  Here we put all of
the ingredients together  and write dynamical equations for an
elasto-plastic material in  two dimensions, preserving a clear picture
of energy balance.

In Section \ref{sec:6} we present some arguments in favor of our 
conjecture of proportionality between the rate of creation and annihilation
of STZs and the rate of energy dissipation. We also discuss some details 
that have been left out so far, but still may be important to  obtain 
quantitative agreement with experiment.

\section{Energy concepts in the STZ theory of plasticity} \label{sec:2}

The basic premise of the STZ theory is that the process of plastic
deformation in an amorphous material is due to non-affine
rearrangements  of its particles in certain regions, that are called
shear transformation zones. 
The original STZ theory simplistically considered all STZs as oriented in 
a single preferred direction. A two-dimensional sample was
subjected to pure shear loading with a principal axis of the
deviatoric stress tensor coinciding with the preferred direction.
Throughout this section we will adhere to the same propositions.

To be specific,  we  will call the zones elongated along  the $y$-axis as
``$+$'' zones  and the zones elongated along the $x$-axis as ``$-$''
zones. We will denote the density of zones in the ``$+$'' state by
$n_+$,  and in the ``$-$'' state   by $n_-$. For pure shear the
deviatoric stress tensor has the form:  $s_{xx}=-s$, $s_{yy}=s$,
$s_{xy}=0$.

Following \cite{FALK:1998}, we can think of the  plastic strain rate
as the result of  transitions between  the states of STZs:
\begin{equation}\label{dotepsilon:1}
\dot{\varepsilon}_{pl}= \lambda v \left(R_{-} n_{-}-R_{+} n_{+} \right)
\; ,
\end{equation}
where $\dot{\varepsilon}_{pl}$ is the $yy$-component of the  plastic
strain rate tensor,  $R_{+}$ is the rate of transitions from ``$+$''
to ``$-$'' states,   $R_{-}$ is the rate of transitions from ``$-$''
to ``$+$'' states, $\lambda$ is the elementary increment of the
shear strain, and  $v$ is a volume of the order of the STZ volume. 
Generally transition rates are functions of stress $s$ or, 
equivalently, of the dimensionless variable $s/\bar{\mu}$,  where
$\bar{\mu}$ can be interpreted as a sensitivity modulus\cite{FALK:1998}. 
This modulus has dimension of stress or energy density.
Equation (\ref{dotepsilon:1}) also implies that all STZs 
have the same size, and therefore the constants $\lambda$ and $v$ are 
the same for all zones.

We suppose that STZs can also be annihilated and created, with the
annihilation rate $R_a$ and creation rate $R_c$. 
The creation rate, unlike transition and annihilation rates,
can be understood only as a quantity defined per unit volume. 
Thus, we  have:
\begin{eqnarray} \label{nplusminus}
\dot{n}_{\pm} = R_{\mp} n_{\mp} - R_{\pm} n_{\pm}  - 
R_{a} n_{\pm} + R_{c} \; .
\end{eqnarray}

We can rewrite Eqs. (\ref{dotepsilon:1}), (\ref{nplusminus}) in a more 
convenient form.  If we introduce a parameter $\tau_0$ that specifies 
some time scale for transitions, and define rate functions 
${\cal S} = \tau_0 (R_--R_+)/2$, ${\cal C}  = \tau_0 (R_-+R_+)/2$, 
${\cal T}={\cal S}/{\cal C}$, $\Gamma=\tau_0 R_a$, 
densities $n_\infty=2 R_c/R_a$, 
$n_{tot}= n_+ + n_-$, $n_\Delta= n_+ - n_-$, and dimensionless quantity
$\epsilon_0 = \lambda v n_\infty$,  we get:
\begin{eqnarray}\label{homo_gen:1}
\tau_0 \dot{n}_\Delta &=&\frac{2  n_\infty \tau_0 }{\epsilon_0}
\dot{\varepsilon}_{pl} - \Gamma n_\Delta \; , \\ \label{homo_gen:2} 
\tau_0 \dot{n}_{tot}&=& \Gamma (n_\infty - n_{tot}) \; ,\\ \label{homo_gen:3} 
\tau_0 \dot{\varepsilon}_{pl}&=&\frac{\epsilon_0{\cal C} }{n_\infty}  \left(
{\cal T}  n_{tot} - n_\Delta \right) \; .
\end{eqnarray}
This system of equations is completely determined if we define
the functions $R_\pm$, $R_a$ and $R_c$, which was
done in \cite{FALK:1998}. In this paper we will postpone choosing
specific forms of the transition  rates and corresponding functions
$\cal C$, $\cal S$, and first focus on the creation and
annihilation rates. 

From (\ref{nplusminus}) we see that an important
feature of this theory is that  creation and annihilation  of STZs are
independent of their orientations and occur  with equal probability for
both  orientations. This is not a completely trivial assumption. We
disregard  the possibility that creation and particularly annihilation
can happen in connection  with transition processes, and thus be more
intense for one orientation  of STZs than the other. However, the
assumption that  the  creation and annihilation rates are independent
of  orientation is simple and plausible. Another observation we can make
is that the creation rate is very likely to depend on the 
structure of material, 
or in other words, on such characteristics  as  
packing fraction, free volume or structural disorder, as this rate 
is not only a dynamical, but also a structural property. 
This is also expressed in the fact 
that we can define it per volume of material, but not per STZ. On the 
other hand, the annihilation rate, as well as the transition rates, is less
likely to depend on the structure of  material. This is expressed in 
the fact that they can be defined as rates per STZ, and can be thought of
as properties of STZs, but not the surrounding material, the influence of 
which on individual STZs can be described by averaged quantities, 
such as average stress. In further discussion we will assume  that 
changes in the structure of material 
can be described by changes in STZ degrees of freedom only.   
Thus, our only internal dynamical variables are $n_{tot}$ and $n_{\Delta}$, 
while $n_\infty$ is assumed to be a  constant.

It was proposed in
\cite{FALK:1998}  to make the rates of creation and annihilation
proportional to the  rate of plastic work $2 s
\dot{\varepsilon}_{pl}$.  A peculiarity of this expression
mentioned earlier is that these rates, by definition always positive
quantities,  can  become negative.  This happens because plastic work
does not entirely dissipate.

In general, the rate of plastic work done on a system can be represented 
in the form
\begin{equation} \label{plasticwork:psiQ}
2 s \dot{\varepsilon}_{pl} = \frac{d \psi}{d t} + {\cal Q} \; ,
\end{equation}
where $\psi$ is the energy that is stored in the plastic degrees  of
freedom and in principle can be recovered,   and $\cal Q$ is the
dissipation rate -- a non-negative  function of  stresses and internal
variables.  

As annihilation and creation rates themselves are non-negative, 
we propose to make them proportional to the rate of 
dissipation ${\cal Q}$. 
We will give some reasons why this proportionality can be true 
in section \ref{sec:6}, but at the moment this proposition 
should be viewed  as a conjecture that provides a physically sensible 
model and adequately describes mechanical and thermodynamical phenomena
in amorphous solids.

Now we are in a position to  derive formulas for  
$\cal Q$,  $R_a$ and  $R_c$. We write ${\cal Q} = A R_a = A \Gamma / \tau_0$,
where $A$ is a coefficient determining the proportion in which 
dissipated energy drives creation and annihilation rates. 
Generally, this coefficient can be a function of total STZ density $n_{tot}$, 
but not $n_\Delta$, meaning that dissipation produces creations and 
annihilations of STZs independently of their average orientation already
present in the sample. Later we will refine our conjecture and postulate that
the annihilation and creation rates are proportional to the rate of energy
dissipation not simply per volume, but per STZ.
Thus, the coefficient
$A$ will be proportional to $n_{tot}$.  As the energy $\psi$ depends  only on the 
internal  variables $n_\Delta$ and $n_{tot}$, we have:
\begin{equation}
\frac{d \psi}{d t}=\frac{\partial \psi}{\partial n_\Delta}
\dot{n}_{\Delta} + \frac{\partial \psi}{\partial n_{tot}} \dot{n}_{tot}
\; .
\end{equation}
Then, using  (\ref{homo_gen:1}), (\ref{homo_gen:2}) and
(\ref{homo_gen:3}), we derive from  (\ref{plasticwork:psiQ}):
\begin{equation}\label{general:Gamma}
\Gamma=\frac{2  \tau_0 \dot{\varepsilon}_{pl}(s - \frac{n_\infty}{\epsilon_0}
\frac{\partial \psi}{\partial n_\Delta})}{ A -  n_\Delta
\frac{\partial \psi}{\partial n_\Delta} + 
(n_\infty-n_{tot} )  \frac{\partial \psi}{\partial n_{tot}} } \; .
\end{equation}
In (\ref{general:Gamma}) we must  choose
$\psi$ in such a way that $\Gamma$ is always non-negative.  If we
look at $\Gamma$ as a function of $s$, we conclude that both the
numerator  and the denominator must always be  positive
independently. The  numerator is guaranteed to be positive if its two
factors always become  zero simultaneously, that is at $s_0 =\bar{\mu} {\cal
T}^{-1} (n_\Delta/n_{tot})$, where ${\cal T}$ is assumed to be monotonic,  
and ${\cal T}^{-1}$ is the inverse
function of ${\cal T}$. This gives:
\begin{equation}\label{dpsisdndelta}
\frac{\partial \psi }{\partial n_\Delta}=  \frac{\epsilon_0  \bar{\mu}}{n_\infty} 
{\cal T}^{-1} (n_\Delta/ n_{tot} ) \; .
\end{equation}
From (\ref{dpsisdndelta}), it follows that $\psi$ as a function of $n_\Delta$ 
is defined uniquely. If we suppose that the energy $\psi$ must be extensive in  
$n_{tot}$,  we get:
\begin{equation}\label{psi:s2}
\psi= \epsilon_0 \bar{\mu} \, \frac{n_{tot}}{n_\infty}  
\left(P\left( \frac{n_\Delta}{n_{tot}} \right)  + \kappa \right)\; ,
\end{equation} 
where $P(\xi)= \int_0^{\xi} {\cal T}^{-1}(x) \, d x$ and  $\kappa$ is
a constant. 
The term proportional to $\kappa$ plays an interesting role here. 
It determines how much energy is stored in the material due to the 
presence of the STZs.
This energy can be recovered if the sample is annealed
and thus the number of STZs is reduced.
However, in the low-temperature theory we do not have any way 
to reduce the density of STZs if it is less than $n_\infty$
(see Eq. (\ref{homo_gen:2})). Therefore, if we are conducting 
mechanical tests only, this part of the energy
appears to be dissipative, although in general it is not.

Now we refine our conjecture and postulate that  the annihilation and 
creation rates are  proportional to the dissipation rate per STZ. 
We can rewrite our equations  in a simpler form by defining 
$\Lambda= n_{tot}/n_{\infty}$, $\Delta=n_\Delta/n_\infty$, 
$A = a \epsilon_0 \bar{\mu} n_{tot} / n_\infty$, 
$\Psi = \psi/ \epsilon_0 \bar\mu$, $\tilde s = s /\bar \mu$. 
Equations (\ref{homo_gen:1}), (\ref{homo_gen:2}), 
(\ref{homo_gen:3}),
(\ref{general:Gamma}) and (\ref{psi:s2}) then give:
\begin{eqnarray}\label{homo:1}
\tau_0 \dot{\Delta}&=&2 {\cal C}(\tilde s)({\cal T}(\tilde s) \Lambda - 
\Delta ) - \Gamma
\Delta \; , \\ \label{homo:2} 
\tau_0 \dot{\Lambda}&=&\Gamma (1 - \Lambda )
\; ,\\ \label{homo:3} 
\tau_0 \dot{\varepsilon}_{pl}&=&\epsilon_0 {\cal C}(\tilde s)
\left(\Lambda {\cal T}(\tilde s) -  \Delta \right) \; , \\ \label{Gamma}
\Gamma &=& \frac{2 {\cal C}(\tilde s) (\Lambda {\cal T}(\tilde s) - \Delta)(\tilde s -
{\cal T}^{-1}  (\Delta / \Lambda))} {M(\Lambda, \Delta)} \; ,\\ \label{Psi:1} 
\Psi &=& \Lambda ( P (\Delta/\Lambda)) + \kappa ) \; ,
\end{eqnarray}
where the denominator of $\Gamma$ is
\begin{equation}\label{M:1}
M(\Lambda, \Delta)= a \Lambda - \frac{\Delta}{\Lambda} {\cal T}^{-1}\left(
\frac{\Delta}{\Lambda} \right) +(1- \Lambda) \left( P\left(
\frac{\Delta}{\Lambda} \right) + \kappa \right) \; .
\end{equation}
In earlier papers \cite{LANGER:2003, FALK:2004}, where we used
the quasilinear approximation, we chose $a=1$ and $\kappa=1/2$. But
these parameters have a physical significance and we will later study
how their choice influences the behavior of material.

\begin{figure}
      \includegraphics[angle=-90,
      width=0.7\columnwidth]{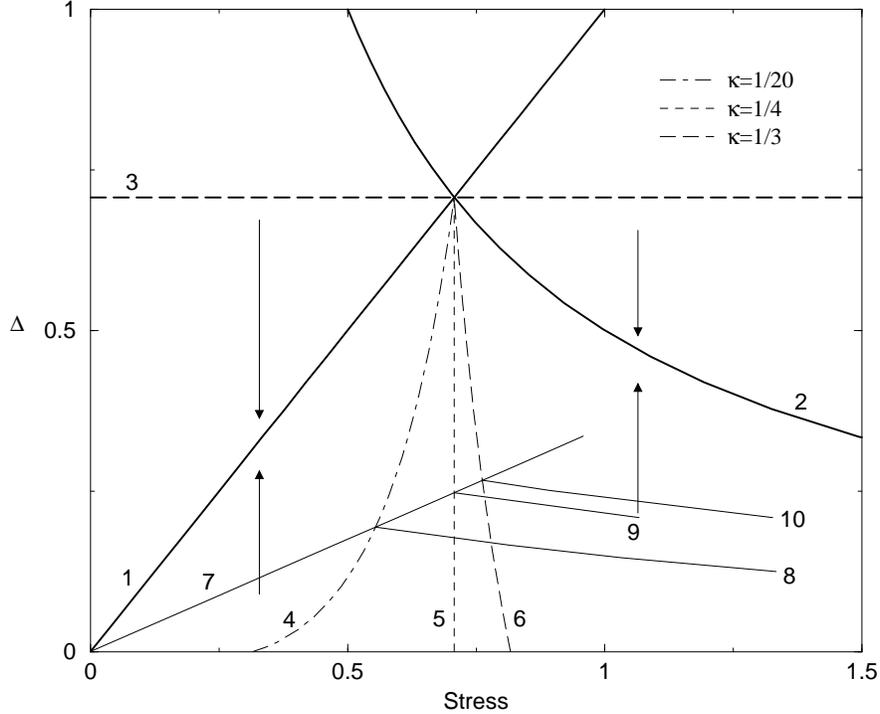} 
\caption{General $\tilde s$-$\Delta$ diagram. Here it is plotted for  
specific parameters of quasilinear model (Sec. \ref{sec:3examples:A}), but 
the diagram's topology is the same for the general case. The thick solid lines 
show two steady states -- 
jammed(1) and flowing(2). The arrows show regions where $\Delta$ increases or 
decreases,
which is determined by the sign of $\dot{\Delta}$ from Eq. (\ref{homo:1}). 
The thick dashed 
line(3) shows the saturation value of $\Delta$, when $\Lambda=1$. The three 
thin dashed and dash-dotted lines(4-6) are the $m_\Lambda$-lines for fixed $a$ and 
the three different values of $\kappa$. The thin solid lines(7-10) show 
quasi-equilibrium branches
for those three values of $\kappa$ at some initial value of $\Lambda$.} 
\label{fig:flowdiagram}
\end{figure}

Let us now  look at the locus of the equilibrium points
$\dot{\Delta}=0$ in the $\tilde s$-$\Delta$ plane (see Fig. \ref{fig:flowdiagram}).
The importance of these
points is due to the fact that they determine  the two states of the
system --  jammed and flowing.  The line $\tilde s={\cal
T}^{-1}(\Delta/\Lambda)$ is  the locus of jammed states,  here
$\dot{\varepsilon}_{pl}=0$.  The other solution,  $\tilde s={\cal
T}^{-1}(\Delta/\Lambda)+ M(\Lambda, \Delta)/\Delta$, is  the locus of flowing
states, where  $\dot{\varepsilon}_{pl}$ is non-zero. 
Note the role that $\Lambda$ is playing here. 
Its equilibrium value is equal to one. 
Accordingly, the lines plotted for $\Lambda \neq 1$ are not 
true equilibrium branches. 
We will call them  quasi-equilibrium, as they change when $\Lambda$ 
relaxes to one.
    
The jammed and flowing branches can intersect only at the point where 
$M(\Lambda, \Delta)=0$. 
The dissipation rate also diverges at  this point. In general, the value
of the variable $m=\Delta/\Lambda$ at this point is a function of $\Lambda$;
we will denote it as $m_\Lambda$. Because of the divergence
in $\Gamma$, the dynamics of Eq. (\ref{homo:1}) is such that 
$m$ is always less than $m_\Lambda$. Thus, the value of $m_\Lambda$ determines
the maximum number of STZs that may flip in one direction; 
we will call it  the saturation point.  

The function $m_\Lambda$ depends on the  parameters $a$ and $\kappa$. Let us 
look at how  their choice influences function's behavior. 
From Eq.~(\ref{M:1}) we  find that when $\Lambda=0$, 
$m_\Lambda=m_0$ is the solution of the equation 
$\kappa=m_0 {\cal T}^{-1}(m_0)-P(m_0)$,
and when $\Lambda=1$, $m_\Lambda=m_1$ is the solution of the equation 
$a=m_1 {\cal T}^{-1}(m_1)$. 
What happens if $a=a_\kappa \equiv m_0 {\cal T}^{-1}(m_0)$, 
so that  $m_1=m_0$?  We can check that in this case  $M(\Lambda, \Delta)$ 
vanishes for any $\Lambda$, if $m=m_0$, 
meaning that $m_{\Lambda} \equiv m_0$.  Thus, we can formulate an important 
property of  Eq.~(\ref{M:1}): for any $\kappa$  there is an $a = a_\kappa $, 
such that  $m_\Lambda$ is independent of $\Lambda$.
The behavior of the function $m_\Lambda$ is also  simple, if  
$a$ differs from $a_\kappa$. We can prove that if $a>a_\kappa$, 
the function $m_\Lambda$ is monotonically increasing, and if $a<a_\kappa$, 
$m_\Lambda$ is  monotonically decreasing. 

To illustrate different choices of parameters $a$ and $\kappa$, in 
Fig.~\ref{fig:flowdiagram}  we  show plots 
of $\Delta=\Lambda m_\Lambda$ as  functions of 
$\tilde s={\cal T}^{-1}(m_\Lambda)$, 
obtained by varying $\Lambda$, for fixed $a$ and three different values
of $\kappa$. We will call such curves $m_\Lambda$-lines; each of them 
is the locus of intersection points of the quasi-equilibrium jammed and flowing
branches, when $\Lambda$ varies. As the value of $a$ is fixed,  
steady state branches coincide for different $\kappa$ when $\Lambda=1$.  
Of the three values of $\kappa$, the intermediate value  
is such that $a_\kappa$ is equal to the given value of $a$. 

In an elasto-plastic material the total strain rate is given by
\begin{equation}\label{dotepstot}
\dot\varepsilon_{tot}=\dot\varepsilon_{pl}+\dot s /2 \mu \; , 
\end{equation}
where $\mu$ is the shear modulus (see also Eq. (\ref{dtot:1}) and the 
discussion thereof). Let us consider solutions of the system 
(\ref{homo:1}-\ref{homo:3},\ref{dotepstot})
at a constant strain rate  $\dot\varepsilon_{tot}$. 
If the strain rate is small, the $m_\Lambda$-lines coincide with 
the dynamical trajectories  in the  regime when
$\Lambda$ is evolving from some initial value $\Lambda_0$ towards unity. 
In other words, the dynamical trajectory in the $\tilde s$-$\Delta$ plane 
first moves along  the quasi-equilibrium jammed branch calculated 
for $\Lambda=\Lambda_0$ (line 7) until the intersection with the 
$m_\Lambda$-line and then moves along the $m_\Lambda$-line 
(for example, along line 4 for the smallest $\kappa$).
For higher strain rates the dynamical trajectories tend to lie to the right 
of the 
quasi-equilibrium jammed branch and the corresponding $m_\Lambda$-line  and 
evolve from zero to some point on the flowing branch at $\Lambda=1$, 
determined by the value of $\dot\varepsilon_{tot}$. The final value 
of $\tilde s$  
can be smaller than intermediate  values, thus producing a stress overshoot. 
As we can see from Fig.~\ref{fig:flowdiagram}, a stress overshoot is more likely 
to happen for large values of  $\kappa$. For small values of $\kappa$ 
the stress 
increase is usually monotonic.    
    
It is hard to find compelling reasons why in a glassy material 
the saturation value  $m_\Lambda$ should be dependent  on $\Lambda$. 
Therefore, we will suppose that for glasses $a=a_\kappa$. Indeed, 
this assumption produces  behavior typical for glasses, such as essential 
strain rate dependence and stress overshoot. Further, we will also study a case 
when $m_\Lambda$ is dependent
on $\Lambda$. This case may be relevant for description of polycrystals, 
clays or soils,
if deformation in such systems is due  mostly to  rearrangement of individual 
crystals or grains, rather than deformation of grains themselves. The difference
from glasses, to which we particularly wish to refer here, is the presence 
of an additional means of energy dissipation due to friction between constituent 
particles, which we will  model with a larger dissipation coefficient, 
that is, with $a > a_\kappa$. 

Now, to make the discussion clear and to put our previous works
into the current more general framework, we  consider the simple case of 
what  we call 
the quasilinear version of the STZ theory. Such an  analysis was presented 
in much detail in \cite{LANGER:2003}, albeit only with $a=a_\kappa$. 

\section{Examples} \label{sec:3examples}
\subsection{Quasilinear theory}\label{sec:3examples:A}

In the quasilinear theory the transition rate functions are supposed
to be linear functions of the shear stress $s$. Namely, we assume that
${\cal C}(\tilde s)=1$, ${\cal S}(\tilde s) = \tilde s$, so that 
${\cal T}(\tilde s) = 
\tilde s$,  ${\cal
T}^{-1}(\xi)=\xi$, $P(\xi)=\xi^2/2$. From (\ref{Psi:1}) we get
\begin{equation}\label{Psi:Q}
\Psi =  \Lambda ( m^2 / 2   + \kappa) \; ,
\end{equation}
where $m=\Delta/\Lambda$.
The expression (\ref{M:1}) for $M$ becomes 
\begin{equation}\label{MQuas}
M(\Lambda,m)=\Lambda (a - \kappa )+\kappa-(\Lambda + 1) m^2 / 2. 
\end{equation}
We find that $m^2_\Lambda=2(a-\kappa+(2 \kappa -a)/(\Lambda +1))$ and 
$a_\kappa = 2 \kappa $.

In \cite{LANGER:2003}, we chose $a = 2 \kappa = 1$, so that $|m_\Lambda|=1$. 
Then Eq.~(\ref{Gamma}) becomes:
\begin{equation}\label{Gamma:quasi}
\Gamma = \frac{4 \Lambda (\Lambda \tilde s -\Delta )^2} {(1 +
\Lambda)(\Lambda^2 - \Delta^2)} \; .
\end{equation}
Using  Eq.~(\ref{Gamma:quasi}) in the dynamic equations
(\ref{homo:1}-\ref{homo:3}) we find that  non-flowing steady states
occur at $\tilde s=\Delta/\Lambda<1$ and  flowing steady states at
$\tilde s=(1+\Lambda)/(2 \Delta)-(1-\Lambda)\Delta/(2 \Lambda^2)>1$.  The
exchange of stability occurs at $\tilde s=1$. This value  can be naturally
associated with the yield stress.

\begin{figure}
      \includegraphics[angle=-90,
      width=0.7\columnwidth]{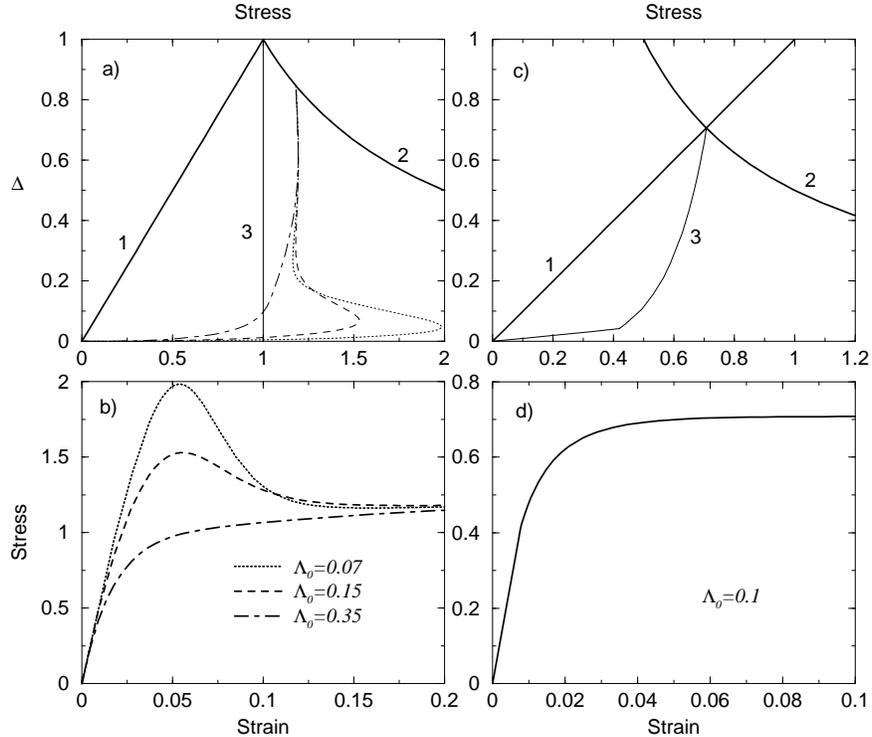} 
\caption{Quasilinear model. $\tilde s$-$\Delta$ and stress-strain diagrams 
for two cases: $m_\Lambda$ is constant (Fig. a, b), 
and $m_\Lambda$ is monotonically increasing (Fig. c, d).  
Lines 1, 2 are the jammed and flowing steady-state 
branches at $\Lambda=1$, line 3 is the $m_\Lambda$-line. Figs. a, c show 
dynamical trajectories for the stress-strain curves 
from Figs. b, d respectively.} 
\label{fig:Quasilinear}
\end{figure}

We solve Eqs.~(\ref{homo:1}-\ref{homo:3}, \ref{dotepstot}) numerically at 
the constant strain rate and show the results in 
Fig.~\ref{fig:Quasilinear} (a, b). We plot 
$\tilde s$-$\Delta$
trajectories and stress-strain curves for three initial values of $\Lambda$. 
When the initial number of STZs is small -- the sample is annealed -- 
a pronounced stress overshoot
is observed. For quenched samples, that is, when the 
initial value of $\Lambda$ is large, the stress 
overshoot disappears.  As shown in \cite{LANGER:2003}, the 
constant strain rate simulations of this model are qualitatively similar
to the  available experimental data \cite{HASAN:1993}.

Now we shall consider other choices of constants $a$ and $\kappa$. 
One artificial
difficulty with the quasilinear approximation is that some choices of 
these constants lead to  $m_\Lambda$ larger than unity, 
thus allowing $m$ to assume non-physical  values. 
To satisfy the condition $|m| \leq 1$,
additional conditions must be imposed on the acceptable values of $a$ 
and $\kappa$. 
There are two regions for parameters $a$ and $\kappa$, where $|m| \leq 1$ 
for all $\Lambda$ :   
(1) when  $\kappa \leq 1/2$ and $\kappa \leq a \leq 2 \kappa$, 
here $a \leq a_\kappa$, and 
(2) when  $ a - \kappa  \leq  1/2 $  and $ 2 \kappa \leq a$, 
here $ a \geq a_\kappa $. 

In Fig. \ref{fig:Quasilinear} (c, d) we plot the results of 
simulation for $a=1/2$, $\kappa=1/20$, that is, when they are in the second
region. The strain rate is small, thus the steady state on 
the $\tilde s$-$\Delta$
diagram almost coincides with the intersection of two steady state branches,
and, as discussed in Sec. \ref{sec:2}, the dynamical trajectory follows along
the quasi-equilibrium jammed branch and then along the $m_\Lambda$-line.  
The stress-strain curve exhibits strain hardening as 
observed  in polycrystals, 
soils or clays \cite{ZHU:2000}. Because such 
a strain-rate curve exists in the limit of an infinitely small strain
rate, it can be rate independent for many decades on the logarithmic scale.
During strain hardening almost all the energy 
goes to creating more STZs. As we already noted, we can not get this 
energy back 
in mechanical tests, so in this sense, such a regime can be considered 
to be  dissipative. 

\subsection{Non-linear STZ model}\label{sec:3b}

The quasilinear model is very useful as a toy model because of its simplicity.
It allows us to proceed much further in analytical and, often, numerical
calculations. But it is mainly useful only to gain  qualitative insight 
into the underlying dynamics, not to look for quantitative predictions. 
The most important  drawback of the quasilinear model 
is its exaggeration of  plasticity at small stresses. This drawback 
can be traced to the form  of function ${\cal C}(\tilde s)$  which 
is constant in the quasilinear approximation, but in reality is vanishingly  
small  at small stresses. This property is also responsible for suppressing 
the dynamics of $\Delta$ at small stresses  and, thus, for memory effects.

The general derivation of section \ref{sec:2} suggests  
that the energetic approach to 
the fully non-linear model will give  qualitatively the same results as those
of  the quasilinear model, while fixing
inaccuracies of the latter. In this subsection we briefly illustrate this
point.
          
In a full STZ model ${\cal C}$ and ${\cal S}$ can be  arbitrary non-linear 
functions of shear stress. The important fact to note is that, unlike
what is assumed in the quasilinear approximation, the 
function  $|{\cal T}(\tilde s)|=|(R_{-}-R_{+})/(R_{+}+R_{-})|$
is always less than unity and asymptotically approaches 
it when  $s \rightarrow \infty$. 
This causes the function $|{\cal T}^{-1}(m)|$  to diverge when $|m|$ 
approaches  unity. Thus, when the denominator (\ref{M:1}) vanishes, the 
value of
$|m|=|\Delta/\Lambda|$ is always less than unity. It can not exceed unity
at any values of parameters $a$ and $\kappa$, but it also cannot be equal to 
one. Value of $m$ equal to one corresponds to the complete saturation --
the case when all STZs are oriented in one direction. It is puzzling
that the non-linear theory does not allow this.   
But  we will see in section \ref{sec:4} that this is what must happen, if
we take into consideration that in amorphous materials the STZs are 
oriented arbitrarily.

Next, to proceed with numerical calculations we  make a  particular
choice of  functions $R_+$ and $R_-$. 
We will assume that functions $R_{\pm}$ have the form offered 
in \cite{FALK:1998}, that is $R_{\pm}= 
\exp \left\{-\beta \exp{(\pm \tilde s)} \right\}/\tau_0 $, where
$\beta=V^*/v_f$, $v_f$ is the  average 
free volume and $V^*$ is of order of the average molecular volume. 
So we find that 
\begin{eqnarray}\nonumber
{\cal C}(\tilde s) &=& \exp{\left (-\beta \cosh \tilde s \right) } 
         \cosh{\left ( \beta \sinh \tilde s \right ) } \; , \\\nonumber
{\cal S}(\tilde s) &=& \exp{\left (-\beta \cosh \tilde s \right) }
         \sinh{\left (\beta \sinh \tilde s \right) } \; ,  \\
{\cal T}(\tilde s) &=& \tanh{\left ( \beta \sinh \tilde s \right)} \; .
\end{eqnarray}

\begin{figure}
      \includegraphics[angle=-90,
      width=0.7\columnwidth]{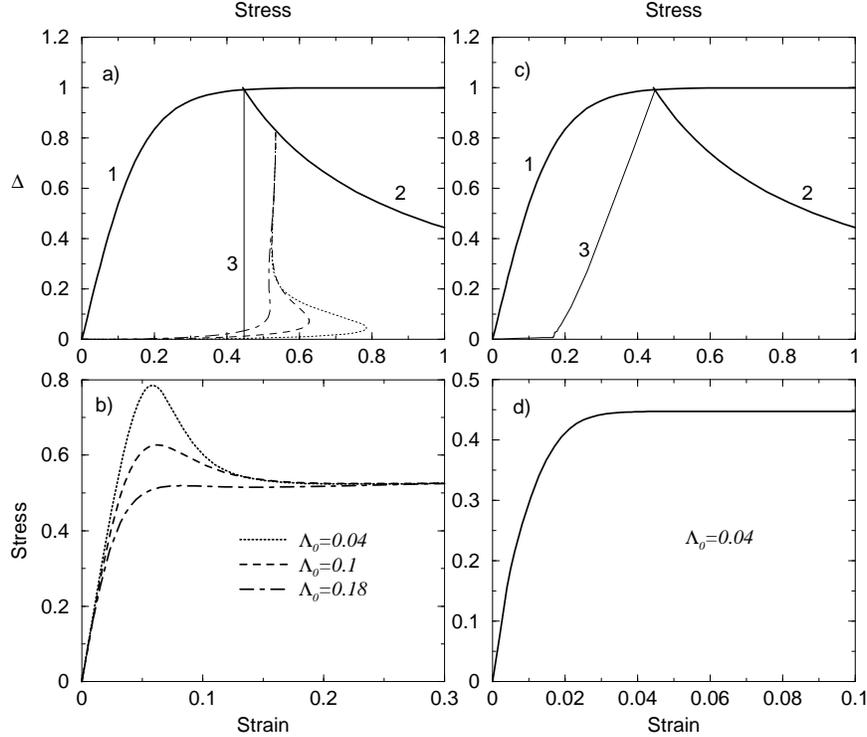} 
\caption{Nonlinear model. $\tilde s$-$\Delta$ and stress-strain diagrams 
for two cases: $m_\Lambda$ is constant (Fig. a, b), 
and $m_\Lambda$ is monotonically increasing (Fig. c, d).  
Lines 1, 2 are the jammed and flowing steady-state 
branches at $\Lambda=1$, line 3 is the $m_\Lambda$-line. Figs. a, c show 
dynamical trajectories for the stress-strain curves from Figs. b, d 
respectively.}  
\label{fig:Nonlinear}
\end{figure}

Now we can find that the function 
${\cal T}^{-1}(m)=\hbox{arcsinh}(\hbox{arctanh}(m) / \beta)$. It diverges
logarithmically at $m \rightarrow \pm 1$,  but the function $P(m)$ and 
consequently the
plastic energy $\Psi$ given by (\ref{Psi:1}) are finite at $|m|=1$.

For the numerical simulations at a constant strain rate loading we used
the parameter $\beta=6$. In Fig. \ref{fig:Nonlinear} (a, b) we have 
$m_\Lambda \equiv 0.992 = const$,
which is obtained with $a=0.444$, $\kappa=1/3$. The stress rate 
is large,  so that the equilibrium point on the flowing branch (2) is far from the 
intersection of the jammed (1) and the flowing (2) lines. As in Fig. 
\ref{fig:Quasilinear} (a, c), we demonstrate
the results for three different values of $\Lambda_0$. In Fig. \ref{fig:Nonlinear} (c, d), 
we chose $a=0.444$, $\kappa=1/15$, so that the $m_\Lambda$-line is monotonically
increasing. The strain rate is chosen to be small so the steady state point 
almost coincides with the intersection of the jammed and flowing branches. 
Note that analytically we can calculate very little in the fully
non-linear model, and even the numerical solution requires not simply solving
the system of differential equations, but also numerically calculating the integral
$P(m)$ at every step. But the analysis from Sec. \ref{sec:2} predicts much 
of the solution's behavior just from knowing how the function $m_\Lambda$ behaves
based on the numerical values of $a$ and $\kappa$. We see that the plots
in Fig. \ref{fig:Nonlinear} are very similar to the plots in Fig. \ref{fig:Quasilinear}
for the quasilinear model, since the $\tilde s$-$\Delta$ diagrams are topologically
the same. 

\section{Isotropic STZ model of plasticity}\label{sec:3}

In this section we generalize the STZ model of plasticity to  the case
of arbitrary spatial orientations of STZs and arbitrary  orientations
of the stress.  The first attempt to make such a generalization
starting  from microscopic basics was made by M. Falk
\cite{FALK-THESIS},  but was not quite complete.

Here again we will consider a two-dimensional homogeneous sample under
 a pure shear.
\begin{figure}
      \includegraphics[angle=0,
      width=0.7\columnwidth]{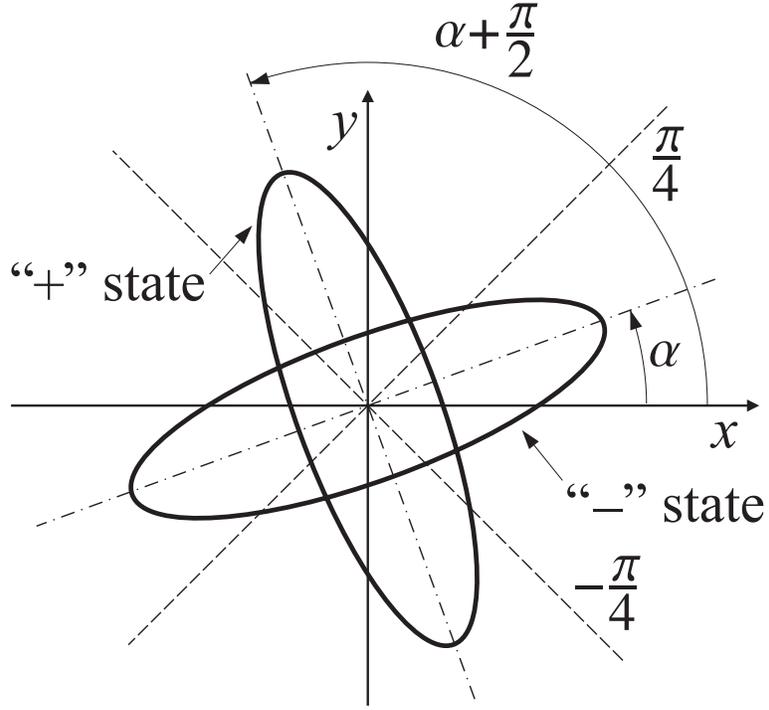} 
\caption{Classification of STZs as being in
``+'' or ``$-$'' states.}  \label{fig:orientation}
\end{figure}
To be specific, we will classify STZs in relation to the  direction of
the $x$ and $y$ axes. We will specify that an STZ is in  the ``$+$''
state if the angle between the direction of  its elongation and  the
$y$-axis is smaller than $\pi/4$;  when the same is true with respect
to the  $x$-axis, we will say that the STZ is in the ``$-$'' state
(see Fig. \ref{fig:orientation}).  Note, we suppose that  the ``$+$''
and ``$-$''  orientations of the zone are perpendicular to each other.
Deviations from a right angle should be described as
fluctuations beyond the mean field theory, and therefore 
will not be considered here.  We write the pure shear in
the form $s_{ij}=\bar{s} d^{\phi}_{ij}$, where $\phi$ is the direction
of the principal axis of the stress tensor, $\bar{s}=\sqrt{s_{ij}
s_{ij}/2}$, and
\begin{equation}
d^{\phi}_{ij}=2 \hat{e}^{\phi}_{i} \hat{e}^\phi_{j} - \delta_{ij}\; .
\end{equation}
In the above equation  $\hat{e}^{\phi}_{i}$ is a unit vector in the
direction $\phi$,  so that $d^{\phi}_{xx}=-d^{\phi}_{yy}=\cos 2 \phi$
and   $d^{\phi}_{xy}=d^{\phi}_{yx}=\sin 2 \phi$. We measure the angle
$\phi$  in the counterclockwise direction  relative to the
$x$-axis. For the purposes of  this  section we could have chosen the
principal axes of the stress tensor to be  oriented along the $x$ and
$y$ axes, but as we will further want to  generalize this discussion
for the case of  arbitrary temporal evolution of the stress, we
suppose   that  $\phi$ is arbitrary.

Then we suppose that only the diagonal component of the shear  stress
tensor in the direction of the zone orientation  (the projection of
the shear stress tensor  on that direction)  is  important for the
dynamics of transitions between the states of this zone.  Thus,  for
the dynamics of the  STZ population  we write:
\begin{eqnarray} \label{nplus2d}
\dot{\chi}_\alpha^{+}=R_{-}(s_\alpha) \chi^{-}_\alpha- R_{+}(s_\alpha)
\chi^{+}_\alpha  -R_{a} \chi_\alpha^{+} + R_{c}\; , \\
\label{nminus2d} \dot{\chi}^{-}_\alpha=R_{+}(s_\alpha)
\chi_\alpha^{+}-R_{-} (s_\alpha) \chi_\alpha^{-} -R_{a}
\chi_\alpha^{-} + R_{c} \; ,
\end{eqnarray}
where $\chi_\alpha^{\pm}$ is the  density of STZs   in the
``$+$''/``$-$'' state oriented at an angle $\alpha$ relative  to the
$x$ axis,  and $s_\alpha=\bar{s}  d^{\phi-\alpha}_{xx}$ is the
projection of the shear stress tensor on the  direction $\alpha$.  At
this moment the density $\chi_\alpha^{\pm}$ is defined for angles
from  $-\pi/4$ to $\pi/4$. Note that all  STZs are included in this
range due to the circular symmetry.

Now we note that our classification of zones as ``$+$'' and ``$-$''
depends on the choice of the direction of $x$ and $y$ axes, which is
arbitrary. If one zone is in the ``$+$'' state in relation to  a
particular direction, then it is in the ``$-$'' state in relation  to
the perpendicular direction, that is,
$\chi^\pm_{\alpha+\pi/2}=\chi^\mp_\alpha$. Our dynamical equations
should not depend on such arbitrariness; they should give  the same
results independently of a reference direction. Therefore,
Eq.~(\ref{nminus2d}) for the angle $\alpha\pm\pi/2$ must be the same
as   Eq.~(\ref{nplus2d}) for the angle $\alpha$, and vice versa. Thus,
we conclude that the following relation for transition rates must hold:
\begin{equation}\label{sym:trrates}
R_+(s_\alpha) = R_-(-s_\alpha)\; .
\end{equation}

We suppose that transitions do not change the volume of
material. Thus,  we must describe the elementary change in strain by a
traceless tensor, which in two  dimensions is proportional to
$d^{\alpha}_{ij}$. Again, we suppose that the magnitude of this
elementary change is always the same, only its  orientation can be
different. In analogy with Section \ref{sec:2} we have:
\begin{equation}\label{dij}
\dot{\varepsilon}_{ij}^{pl}=\lambda v \int_{-\pi/4}^{\pi/4}
 d^{\alpha}_{ij}\left(  R_{+} (s_\alpha) \chi^{+}_\alpha - R_{-}
 (s_\alpha) \chi^{-}_\alpha \right) d \alpha \; .
\end{equation}
The region  of integration in  (\ref{dij}) is chosen  to count every
STZ only once. However, from the symmetry for the angles $\alpha$ and 
$\alpha+\pi/2$ we conclude that the same integral is correct with any 
limits of integration  in the form $[-\pi/4+\gamma ,\pi/4 +
\gamma]$.
 
As in Section \ref{sec:2}, we can introduce rate functions ${\cal S}=
\tau_0 (R_--R_+)/2$, ${\cal C}= \tau_0 (R_++R_-)/2$,  
${\cal T}={\cal S}/{\cal C}$, $\Gamma=\tau_0 R_a$
and also densities $\chi_\alpha^{tot}=\chi^+_\alpha +\chi^-_\alpha$,
$\chi_\alpha^\Delta=\chi^+_\alpha -\chi^-_\alpha$, $\chi_\infty \equiv 
n_\infty  = 2 R_c /R_a$, where, as earlier, tilde means
stress rescaled by $\bar \mu$, that is $\tilde s_\alpha =s_\alpha/\bar \mu$,
$\tilde s_{ij} =s_{ij}/\bar \mu$, $\tilde{\bar{s}}  =\bar s /\bar \mu$.
These functions also have symmetry properties: 
${\cal S}(\tilde{s}_\alpha)= -{\cal
S}(\tilde{s}_{\alpha+\pi/2})$, ${\cal C}(\tilde{s}_\alpha) ={\cal
C}(\tilde{s}_{\alpha+\pi/2})$, ${\cal T}(\tilde{s}_\alpha)= -{\cal
T}(\tilde{s}_{\alpha+\pi/2})$, and
$\chi^{tot}_\alpha=\chi^{tot}_{\alpha+\pi/2}$,
$\chi^\Delta_\alpha=-\chi^\Delta_{\alpha+\pi/2}$. Using these
variables we can rewrite Eqs. (\ref{nplus2d}), (\ref{nminus2d})  and
(\ref{dij}) as:
\begin{eqnarray}\label{dij:2}
\tau_0 \dot{\varepsilon}_{ij}^{pl}&=&\frac{\epsilon_0}{\chi_\infty}  
\int_{-\pi/4}^{\pi/4}
d^{\alpha}_{ij} {\cal C} (\tilde s_\alpha) \left(  {\cal T} (\tilde s_\alpha)
\chi^{tot}_\alpha -  
\chi^{\Delta}_\alpha \right) d \alpha \; ,\\\label{rhodelta} 
\tau_0 \dot{\chi}_\alpha^{\Delta}&=&2 {\cal C}(\tilde s_\alpha)
\left( {\cal T}(\tilde s_\alpha) \chi^{tot}_\alpha- \chi^{\Delta}_\alpha
\right) - \Gamma \chi_\alpha^\Delta \; , \\ \label{rhotot}
\tau_0 \dot{\chi}^{tot}_\alpha
 &=& \Gamma (\chi_\infty  - \chi^{tot}_\alpha) \; .
\end{eqnarray}

These equations are analogs of Eqs.  (\ref{homo_gen:1}),
(\ref{homo_gen:2}),  (\ref{homo_gen:3}), but with arbitrary
orientations of STZs. Instead of the number of STZs in two different
states  their variables  are the densities of STZs with different
orientations.  As it is hard to deal with such equations,  where
$\chi_\alpha$ essentially plays the role of a distribution function,
we further show that these equations can be simplified under
sufficiently relaxed assumptions, and instead of the density
$\chi_\alpha$ we can  introduce its moments -- scalar and tensor
variables.

Instead of the angular density $\chi^{tot}_\alpha$ we can introduce
the total  density of zones in a sample  $n_{tot}=(2 / \pi)
\int_{-\pi/4}^{\pi/4} \chi^{tot}_\alpha d \alpha$, the  equation  for
which is easy to get by integrating  Eq. (\ref{rhotot}). Instead of
$\chi^{\Delta}_\alpha$ we introduce the tensor
$n_{ij}=\int_{-\pi/4}^{\pi/4} d_{ij}^{\alpha}  \chi^\Delta_{\alpha} d
\alpha$. To get dynamical equations for  $n_{ij}^\alpha$ we multiply
Eq. (\ref{rhodelta}) by $d_{ij}^{\alpha}$ and integrate it over
$\alpha$:
\begin{equation} \label{nij}
\tau_0 \dot{n}_{ij} =  2  \int_{-\pi/4}^{\pi/4} d^\alpha_{ij} {\cal
C}(\tilde s_\alpha) \left( {\cal T}(\tilde s_\alpha) \chi^{tot}_\alpha-
\chi^{\Delta}_\alpha \right) d \alpha - \Gamma n_{ij} \; .
\end{equation}

An assumption we will make here is that initially $\chi^{tot}_\alpha$
does not depend on $\alpha$. Then according to Eq. (\ref{rhotot})
$\chi^{tot}_\alpha$ is independent of $\alpha$  at all later
times. Next, in the integral  (\ref{nij}) we  will approximate the
function  ${\cal C}(\tilde s_\alpha)$ by a function  
$\bar{\cal C}(\tilde{\bar{s}})$
that depends not on the projection of the shear stress tensor on a
given direction,  but on the principal value of the shear stress
$\bar{s}$. The only role that the function $\cal C$ played in the
original  paper  \cite{FALK:1998} was to be responsible for memory
effects.  It was a vanishingly small function for small stresses, and
thus   effectively froze the internal variables in an unloaded sample,
preserving information about the previous loading.  Our approximation
keeps such dynamics intact.  Now the integral in Eq. (\ref{nij}) can
be calculated. Together with Eqs.  (\ref{dij:2}), (\ref{rhotot}) our
system becomes:
\begin{eqnarray} \label{dij:3}
\tau_0 \dot{\varepsilon}_{ij}^{pl}&=& \frac{\epsilon_0}{n_\infty} 
\bar{\cal C}(\tilde{\bar{s}})
 \left(\bar{\cal T}(\tilde{\bar{s}})\frac{s_{ij}} {\bar{s}} 
n_{tot} - n_{ij}\right) \; , \\ \label{nij:3} 
\tau_0 \dot{n}_{ij}&=&  2 \bar{\cal
 C}(\tilde{\bar{s}})  \left( \bar{\cal T}(\tilde{\bar{s}}) 
\frac{s_{ij}}{\bar{s}}
 n_{tot} - n_{ij}\right)  - \Gamma n_{ij}   \; , \\ \label{ntot:3}
 \tau_0 \dot{n}_{tot}&=& \Gamma (n_\infty -  n_{tot}) \; ,
\end{eqnarray}
where we denoted
\begin{equation}\label{Tbar}
\bar{\cal T}(\tilde{\bar{s}}) = \int_{-\pi/4}^{\pi/4}d \theta {\cal T} (
\tilde{\bar{s}} \cos 2 \theta) \cos 2 \theta \; .
\end{equation}
In  the derivation of this  system we used the previously discussed
property  that we can change the region of integration to any
quadrant. In more detail the integrals in  (\ref{dij:2}) and  
(\ref{nij}) had been calculated as follows:
\begin{eqnarray} \nonumber
\int_{-\pi/4}^{\pi/4} d_{ij}^{\alpha} {\cal T}(\tilde s_\alpha) d \alpha  &=& 
\int_{-\pi/4}^{\pi/4} d_{ij}^{\alpha} {\cal T}(\tilde{\bar{s}}  \cos 2 (\phi -
\alpha))  d \alpha = \\  \nonumber  &=& \int_{-\pi/4}^{\pi/4}
d_{ij}^{\phi-\theta} {\cal T} (\tilde{\bar{s}} \cos 2 \theta )  d \theta =
d_{ij}^{\phi} \int_{-\pi/4}^{\pi/4}  \cos 2 \theta \,
{\cal T} (\tilde{\bar{s}} \cos 2 \theta ) d \theta \;,
\end{eqnarray}
where $d_{ij}^{\phi}$ is equal to  $ s_{ij}/ \bar{s} $.

Equations (\ref{dij:3}), (\ref{nij:3}) and (\ref{ntot:3}) give the description
of plasticity in the isotropic generalization of the STZ theory.

\section{The proportionality hypothesis for the isotropic STZ model}\label{sec:4}

We now show how to expand the results of Section \ref{sec:2} for  the
isotropic  case. Again, as we will want to generalize results of this
section for the case of arbitrary temporal evolution of the stress, we
suppose  that the principal axes of the tensor $s_{ij}$ do not
necessarily  coincide with the principal axes of the tensor
$n_{ij}$. We write the plastic work done on a system as:
\begin{equation}\label{energybalance:2d}
\dot{\varepsilon}^{pl}_{ij} s_{ij} \equiv \frac{\epsilon_0}{\tau_0 n_\infty} 
s_{ij}  \bar{\cal C}(\tilde{\bar{s}})  \left(\bar{\cal T}(\tilde{\bar{s}})
\frac{ s_{ij}} { \bar{s} } n_{tot} - n_{ij}\right) = 
\frac{d \psi(n_{ij}, n_{tot})}{d t} + {\cal Q}  \; .
\end{equation} 

We will denote $s_{xx}=-s_{yy}=s$, $s_{xy}=s_{yx}=\tau$,
$n_{xx}=-n_{yy}= n_\Delta$,  $n_{xy}=n_{yx}=n_\delta$, and the
invariant of the $n_{ij}$ tensor as
$\bar{n}=(n_\Delta^2+n_\delta^2)^{1/2}$. The energy $\psi$ is now  a
function of three variables, so:
\begin{equation}
\frac{d \psi}{d t}=\frac{\partial  \psi}{\partial n_\Delta } \frac{d
n_\Delta}{d t} + \frac{\partial  \psi}{\partial n_\delta } \frac{d
n_\delta}{d t} +  \frac{\partial  \psi}{\partial  n_{tot}} \frac{d
n_{tot}}{d t} \; .
\end{equation}
As in Section \ref{sec:2}, we suppose that ${\cal Q} = a \epsilon_0 
\bar \mu n_{tot} \Gamma / \tau_0 n_\infty$. 
Writing (\ref{energybalance:2d}) in components and then assuming  that 
the energy $\psi$ can depend on $n_\Delta$ and  $n_\delta$ only 
through $\bar{n}$, we find:
\begin{equation}\label{Q:nonlinear2d}
{\Gamma}= 2 \bar{\cal C}(\tilde{\bar{s}}) \frac{ \left(n_{tot} \bar{\cal
T}(\tilde{\bar{s}})\frac{ s }{ \bar{s} }   - n_\Delta \right) 
\left(
\frac{\epsilon_0}{n_\infty} \tilde{s} -
\frac{\partial \psi}{\partial \bar{n}} \frac{n_\Delta}{\bar{n}}\right)
+ \left(n_{tot} \bar{\cal T}(\tilde{\bar{s}})\frac{ \tau}
{ \bar{s} }  - n_\delta
\right) \left(\frac{\epsilon_0}{n_\infty} \tilde\tau - \frac{\partial \psi}{\partial \bar{n}}
\frac{n_\delta}{\bar{n}}\right)} {a \epsilon_0 \bar \mu 
\frac{n_{tot}}{n_\infty}- \bar{n} \frac{\partial
\psi}{\partial \bar{n}} + 
(n_\infty - n_{tot})\frac{\partial \psi}{\partial n_{tot}} } \; .
\end{equation} 
The rate function $\Gamma=\tau_0 R_a$ must always be positive.  In analogy
with Section \ref{sec:2}, considering  this expression  as a function
of stresses allows us to  conclude that the numerator and the
denominator of $\Gamma$  must always be positive  separately.  For
fixed $n_{\Delta}$, $n_{\delta}$ and $n_{tot}$ and varying $s$, $\tau$
we want the numerator
to pass through zero at a single point $(s_0, \tau_0)$ and be positive 
elsewhere. The numerator becomes equal to zero when its first and 
third brackets are
equal to zero. 
This happens for stresses $s_0=n_\Delta \bar{s}_0 /(n_{tot} \bar{\cal
T}(\tilde{\bar{s}}_0) )$ and $\tau_0=n_\delta \bar{s}_0 /(n_{tot} \bar{\cal
T}(\tilde{\bar{s}}_0) )$.  Now we can express $s_0$ and $\tau_0$ as functions
of the variables $n_\Delta$,  $n_\delta$ and $n_{tot}$ only. 
Noting that  $\bar{s}_0 =(s_0^2+\tau_0^2)^{1/2}$,  we find 
that  $\bar{s}_0=\bar{\mu} \bar{\cal T}^{-1}(\bar{n}/n_{tot})$. Substituting
$\bar{s}_0$ in the expressions for $s_0$ and $\tau_0$,
we get   
$s_0=\bar\mu \bar{\cal T}^{-1}(\bar{n}/n_{tot}) n_\Delta / \bar{n}$,  
$\tau_0=\bar\mu \bar{\cal T}^{-1}(\bar{n}/n_{tot}) n_\delta / \bar{n}$.

If the second and the fourth brackets also pass through zero at this
point,  they will  always have the same sign as the first and the
third brackets  correspondingly,  ensuring positiveness of the
numerator. Thus,  from either the second or the fourth bracket we find:
\begin{equation}
\frac{\partial \psi}{\partial \bar{n}}=\frac{\epsilon_0 \bar\mu}{n_\infty}  
\bar{\cal T}^{-1}\left(\frac{\bar{n}}{n_{tot}}\right) \; .
\end{equation}
Therefore, we find that the energy $\psi$ has the same form as  
in Section \ref{sec:2} 
Eq. (\ref{psi:s2}):
\begin{equation}\label{psi:s4}
\psi=\epsilon_0 \bar\mu \frac{n_{tot}}{n_\infty} 
\left(P\left(\frac{\bar{n}}{n_{tot}}\right)+\kappa \right) \; .
\end{equation}

Now we can write out our final result for the tensorial generalization of the 
low temperature STZ theory of plasticity in the form analogous to 
Eqs. (\ref{homo:1}-\ref{M:1}). 
If again we denote  $\Lambda=n_{tot}/n_{\infty}$,
$\Delta_{ij}=n_{ij}/n_\infty$, $\bar{\Delta}=\bar{n}/n_\infty$,  we get:
\begin{eqnarray} \label{dij:4}
\tau_0 \dot{\varepsilon}_{ij}^{pl}&=& \epsilon_0  
\bar{\cal C}(\tilde{\bar{s}})
 \left( \Lambda \bar{\cal T}(\tilde{\bar{s}})\frac{s_{ij}} {\bar{s}}  -
 \Delta_{ij}\right) \; , \\ \label{deltaij:4} 
\tau_0 \dot{\Delta}_{ij}&=&  2
 \bar{\cal C}(\tilde{\bar{s}})  \left( \Lambda \bar{\cal T}(\tilde{\bar{s}})
 \frac{s_{ij}}{\bar{s}}  -  \Delta_{ij}\right)  - \Gamma \Delta_{ij}
 \; , \\ \label{lambda:4} 
\tau_0 \dot{\Lambda}&=&\Gamma (1 - \Lambda) \; , \\
 \label{Gamma:4} 
{\Gamma} &=&   \frac{\bar{\cal C}(\tilde{\bar{s}}) (\Lambda
 \bar{\cal T}(\tilde{\bar{s}}) s_{ij} / \bar{s}  -  \Delta_{ij})(\tilde{s}_{ij} - \bar{\cal T}^{-1}(\bar{\Delta} / \Lambda) \Delta_{ij} / \bar{\Delta}
 )}  {M(\Lambda, \bar{\Delta} )} \; .
\end{eqnarray}
Expressions $M(\Lambda, \bar \Delta)$ and $\Psi(\Lambda, \bar \Delta)=\psi/ \epsilon_0 \bar\mu$ are the same as  (\ref{M:1}), (\ref{Psi:1}) with $\Delta$ replaced everywhere by $\bar\Delta$. 

Finally, we can compare our tensorial theory with arbitrary spatial 
orientations of STZs and arbitrary loading, derived in this section, 
with the limited STZ theory of Section \ref{sec:2}
for STZs oriented only along two preferred axes and pure shear loading. 
If we consider pure shear in the generalized STZ theory of this section,
we must assume that the principal axes of tensors $s_{ij}$ and $\Delta_{ij}$ 
are the same. 
Thus, for pure shear  Eqs. (\ref{dij:4}-\ref{lambda:4}) 
become the same as Eqs. (\ref{homo:1}-\ref{homo:3}).
Therefore, the results of Section \ref{sec:2} hold for the 
STZ theory generalized here.
We also note that in the discussion of Section \ref{sec:2} an important 
role was
played by the saturation point -- the value of $m$ for which  
no further transitions were possible.  
In the isotropic case, when all STZs are switched in one direction, 
that is, when  $|\chi_\alpha^\Delta|=\chi^{tot}$,
we can find from the expressions for  $n_{tot}$ and $n_{ij}$ of 
Sec. \ref{sec:3} that $\bar{n}=n_{tot}$. 
For other orientational distributions, 
when  $|\chi_\alpha^\Delta| < \chi^{tot}$ at least in some 
interval of angles,  we have  
$m =\bar{n}/n_{tot} <  1$. 
However, the projection of the stress tensor on the directions in 
the narrow strips under angles
$\pm \pi/4$ to the principal axes of the stress tensor is
small, for any finite value of $\tilde{\bar s}$, leading to 
$|\chi_\alpha^\Delta| < \chi^{tot}$ at least for those angles.
Therefore, we must expect that the saturation point will be reached 
at $m = \bar{n}/n_{tot} < 1$, as has been assumed in Sec. \ref{sec:3b}.   

\section{Continuum equations and energy balance}\label{sec:5}

The plasticity described by the STZ theory can be incorporated  into a
continuum theory that describes elastic and plastic behavior of
viscoelastic solids  using a general framework, discussed, for
example, in \cite{MALVERN, LUBLINER}.

\subsection{The STZ theory of plasticity in a spatially 
inhomogeneous situation}

We start with the generalization of the isotropic STZ model of
plasticity for a spatially inhomogeneous situation.

To make the physical picture clear,  we will now discuss details  omitted
for simplicity in the previous sections.  Let us consider a small
region of material, much smaller  than the size of the sample, but
much larger than individual atoms and  inter-atomic distances. This
region contains many STZs of all possible  orientations, but from a
macroscopic point of view it is infinitesimally small and is
identified by its coordinates only.  Thus, we are on a mesoscopic
scale.

As this region contains many STZs, we consider   the {\em average}
effect of transitions between their states  (which are changes in the
positions  of atoms on  the microscopic level)  on this region as a
small part of the sample. From this point of view the  transition
between  the states of  an STZ gives rise to a  change  of strain at
the point where this region is.

Further we will describe the material by what is called  the
referential description \cite{MALVERN}. Namely, suppose  that we are
sitting in  the material coordinate system and then at some time $t$
we freeze our  frame of reference and describe the evolution of the
material  during an infinitesimally small time interval in this frozen
frame of  reference. We can see that the discussion of Section \ref{sec:3} is
correct even  for an inhomogeneous situation in the material frame of
reference,  when the coordinate system not only moves with the
particular small  region of material, but also rotates with it.  In
the referential frame of reference we must exclude the effect of
translational and rotational motion  to make sure that we consider
the same region of material under the same angle.

Thus, instead of the time derivative of angle dependent quantities
$\chi^{\pm}_{\alpha}$, $\chi^{\Delta}_{\alpha}$,
$\chi^{tot}_{\alpha}$, the dot in the expressions (\ref{nplus2d}),
(\ref{nminus2d}) and later must denote a complete co-rotational
derivative:
\begin{equation}
\dot{\chi}_\alpha \equiv  \frac{\partial \chi_\alpha}{\partial t}+
v_i \frac{\partial \chi_\alpha}{\partial x_i}+ \omega \frac{\partial
\chi_\alpha}{\partial \alpha} \; ,
\end{equation}
where $v_i$ and $\omega$ are the translational velocity and the
angular speed of our region.  When deriving Eq. (\ref{nij}) the integral
$\int^{\pi/4}_{-\pi/4} d^{\alpha}_{ij}\dot{\chi}^\Delta_\alpha d \alpha$
gives the tensorial co-rotational derivative 
\begin{equation}
\frac{{\cal D} n_{ij}}{{\cal D} t} \equiv \frac{\partial
n_{ij}}{\partial t}+ v_k \frac{\partial n_{ij}}{\partial x_k} + n_{ik}
w_{kj}-w_{ik} n_{kj} \; ,
\end{equation}
where $w_{ij}=1/2(\partial v_i/ \partial x_j - \partial v_j/ \partial
x_i)$ denotes the spin tensor. This co-rotational derivative must be used 
in place of $\dot n_{ij}$ in (\ref{nij}) and further.
Correspondingly, instead of the time derivative  $\dot n_{tot}$ in the
expression (\ref{ntot:3}) we get  the total derivative
\begin{equation}
\dtotal{ n_{tot}}{t}=\dpartial{n_{tot} }{t}+ v_i \dpartial{n_{tot}
}{x_i} \; ,
\end{equation}
as the rotational part integrates out.
Finally, in the referential frame of reference the time  derivative of
the small strain tensor $\dot{\varepsilon}^{pl}_{ij}$  is equal  to
the rate of deformation tensor $D^{pl}_{ij}$.

Considering the above,  the system (\ref{dij:3}-\ref{ntot:3}) becomes:
\begin{eqnarray} \label{dij:5}
\tau_0 D_{ij}^{pl}&=& \epsilon_0 f_{\epsilon}(\rho_0/\rho)  \bar{\cal C}(\tilde{\bar{s}})
 \left(\bar{\cal T}(\tilde{\bar{s}})\frac{s_{ij}} {\bar{s}} n_{tot} -
 n_{ij}\right) \; , \\ \label{nij:5} 
\tau_0 \frac{{\cal D} n_{ij}}{{\cal D}
 t}&=&  2 \bar{\cal C}(\tilde{\bar{s}})  \left( \bar{\cal T}(\tilde{\bar{s}})
 \frac{s_{ij}}{\bar{s}} n_{tot} - n_{ij}\right)  - \Gamma n_{ij}   \; ,
 \\ \label{ntot:5} 
\tau_0 \dtotal{{n}_{tot}}{t}&=&\Gamma(n_\infty  -  n_{tot}) \; .
\end{eqnarray} 

In (\ref{dij:5}) we took into account that the  elementary strain,
which is  due to a transition between STZ states, can depend on the
local density of material $\rho$ ($\rho_0$ denotes  some reference
density).  We will discuss this point later.

\subsection{Continuum theory of elasto-plastic deformation}

Here we write out a complete set of equations needed to describe
arbitrary elasto-plastic deformation of material.  We also make
an effort to demonstrate the energy balance properties  of our system of
equations.
This question is certainly not new for a system with constitutive
relations in the rate form. However, we consider it important to show
how plasticity described by the STZ theory can be incorporated 
into such a framework.

First, we need to assert that our set of equations contains general
equations which are true for any material -- the conservation of mass
and  momentum balance  equations:
\begin{eqnarray} \label{eq:mass:1}
&&\frac{d \rho}{d t} + \rho \frac{\partial v_i}{\partial x_i}=0 \; ,
 \\  \label{Gbalance} &&\rho a_i = \frac{\partial
 \sigma_{ij}}{\partial x_j} \; ,
\end{eqnarray}
where $a_i$ is the acceleration of material points,  which in an
inertial coordinate system is equal to $d v_i /d t$, and $\sigma_{ij}$
is the true stress.

We describe the material properties by a set of constitutive
equations,  which  also includes equations for internal variables.  To
describe a viscoelastic solid,  we additively decompose the total
strain rate tensor $D^{tot}_{ij}=1/2(\partial v_i/\partial x_j +
\partial v_j/\partial x_i )$ as the  sum of elastic and plastic parts,
which is  true under the assumption that elastic strain is small:
\begin{equation}\label{dtot:1}
D^{tot}_{ij}=D^{el}_{ij} + D_{ij}^{pl} \; .
\end{equation}

We would like to describe elastic behavior of the material simply  by
Hooke's law, but since  here we are dealing with large deformations of
solids and our equations are in the rate form, we need to take into
account  at least to some extent the dependence of the elastic
properties of the  material  on its density. As we will see, this is
dictated by the  conservation of elastic energy. It is convenient  to
postulate that the equation of state of  the material is defined by
a function $F_K$:
\begin{equation}\label{eq:state}
p= - K f_K (\rho_0/\rho) F_K(\rho_0/\rho)  \; ,
\end{equation}
such that $f_K(x)=F^\prime_K(x)$, $F_K(1)=0$, $f_K(1)=1$.  In the
above equation $p$ is the true pressure and $\rho_0$ is the reference
density of the material, which is convenient (but not necessary  for
further discussion) to assume to be the density of the material at
zero pressure.  The spherical part of the elastic response is fully
described by this  equation and,  in fact, $K$ here is the bulk
modulus. Now we can introduce  the conjugate stress and strain
measures (the strain measure is given implicitly, by defining  only
the rate of deformation):
\begin{equation}
\tilde{p}=p/f_K(\rho_0/\rho)\; ; \; \; \; \tilde{D}_{ii}=D_{ii}
 f_K(\rho_0/\rho) \rho_0/\rho  \; .
\end{equation}
Then, according to (\ref{dtot:1}) and (\ref{eq:mass:1}), we can  write
the rate  form  of Hooke's law as
\begin{equation}
\tilde{D}^{el}_{ii}=-\frac{1}{K}\frac{d \tilde{p}}{d t} \; ,
\end{equation}
which coincides with the  usual form in the case of small
deformations.  Similarly, 
for the deviatoric part of elastic response we have:
\begin{equation}\label{Hooke's}
(\tilde{D}^{el}_{ij})^{dev}=\frac{1}{2 \mu}\frac{{\cal D}
\tilde{\sigma}_{ij}^{dev}}{{\cal D} t} \; ,
\end{equation}
where the conjugate  stress  and strain measures are:
\begin{equation}
\tilde{\sigma}_{ij}^{dev}= \sigma_{ij}^{dev}  / f_\mu(\rho_0/\rho)\; ;
\; \; \;  (\tilde{D}^{el}_{ij})^{dev}=(D^{el}_{ij})^{dev}
f_\mu(\rho_0/\rho)   \rho_0/\rho \; .
\end{equation}
The conservation of mass equation (\ref{eq:mass:1}), the momentum
balance equation (\ref{Gbalance}), the constitutive equations
(\ref{dij:5}), (\ref{dtot:1}), (\ref{Hooke's}), 
the equation of state (\ref{eq:state}), 
and the equations
for  dynamics of internal variables (\ref{nij:5}),  (\ref{ntot:5})
constitute a  full  system of equations, which describe elasto-plastic
behavior of a material.  Those equations possess the property of frame
indifference  \cite{OLDBROYD:1950, MALVERN, LUBLINER}.  In particular,
we used this system in a simplified form  for simulations of necking
\cite{EASTGATE:2003}.

The energy balance equation can be derived from the momentum  balance
equation.  This derivation is very well known for the
balance of energy in a volume fixed in  space, but is
less known for the case  we are interested in here, when 
the balance of energy is considered in the  volume of material. By multiplying
Eq. (\ref{Gbalance}) by $v_i$ and  integrating  over some  arbitrary
material volume $V$ we get:
\begin{eqnarray}    
\int_{(V)} \left( \rho\frac{d}{d t} \frac{v^2_i}{2}+
\frac{\rho}{\rho_0} \frac{d}{d t}  \frac{\tilde{p}^2}{2 K} +
\frac{\rho}{\rho_0}\frac{d}{d t}
\frac{(\tilde{\sigma}^{dev}_{ij})^2}{4 \mu}+ D^{pl}_{ij}
\sigma_{ij}\right) d V =  \int_{(S)} v_i \sigma_{ij} d S_j \; .
\end{eqnarray} 

The factor $\rho/\rho_0$ in front of the total derivative plays an
important role.  Without it we would not be able to move
differentiation over time in front of the integral. But as
$\rho/\rho_0$ is the  Jacobian of the transformation from the
coordinate system $x_i (t)$ to the  reference state $x_i (0)$, we can
first  change the variable of integration to $x_i(0)$, then put the
time derivative  in front of the  integral (instead of a total
derivative we will only be left with a derivative over time), and
finally we can change variables of integration back to $x_i(t)$. This
is a purely mathematical procedure. It can be physically  interpreted
in the following way: instead of integrating over the time  varying
volume $dV$,  we integrate over the conserved mass $\rho dV$.   We get:
\begin{eqnarray} \label{energy_balance}
\frac{d}{d t} \int_{(V)} \left(\frac{1}{2}\rho v^2_i+
\frac{\rho}{\rho_0}  \frac{\tilde{p}^2}{2 K} +
\frac{\rho}{\rho_0}\frac{(\tilde{\sigma}^{dev}_{ij})^2}{4 \mu}+
\frac{\rho}{\rho_0}\psi \right) d V = \\ \nonumber \int_{(S)} v_i
\sigma_{ij} d S_j - \int_{(V)} {\frac{\rho}{\rho_0}\cal Q}  d V \; .
\end{eqnarray}
Above we also supposed that the plastic work can be expressed as
\begin{equation}\label{form_pl_en}
D^{pl}_{ij} \sigma_{ij}=\frac{\rho}{\rho_0} \left(  \frac{d
\psi(n_{ij}, n_{tot})}{  d t} + {\cal Q} \right).
\end{equation}
Equation (\ref{energy_balance}) shows that energy in a particular
volume of material consists of kinetic,  elastic and plastic parts. It
is changed by the work of external forces and it
also dissipates  due to plastic processes.

An important example relevant to above discussion is the 
Kirchoff stress tensor
$\tilde{\sigma}_{ij}=\sigma_{ij} \rho_0 / \rho$, which is often used
in engineering applications\cite{MCMEEKING:1975} and standard engineering
software \cite{ABAQUS}. This stress tensor 
is conjugate to the rate of deformation
tensor $D_{ij}$ \cite{HILL:1959}.  We get such  a formulation, if we
set $F_K(x)=\ln x$, $f_K=1/x$, $f_\mu=1/x$.  This formulation assumes
the following equation of state:  $p=K (\rho/ \rho_0) \ln{\rho/\rho_0}$.

Now we return to the assumption (\ref{form_pl_en}). For
this equation to be valid, the   plastic rate of deformation tensor
must be dependent on the density of material.  This dependency 
has already been  explicitly introduced in (\ref{dij:5}).  
At this point it is convenient  to generalize our
description of plasticity and also take into account the possible  
dependence of transition rates on the local density of material, which
we include in the definition of the stress tensor $s_{ij}$:
\begin{equation}
s_{ij}=\sigma_{ij}^{dev} f_\epsilon (\rho_0/\rho ) \rho_0 /\rho \; .
\end{equation} 
Then the density of the rate of plastic work  $D_{ij}^{pl}
\sigma_{ij}$ can be expressed as  a product of $\rho/\rho_0$ and  a
function of $s_{ij}$, $n_{ij}$, $n_{tot}$,  but not density. The
equation (\ref{form_pl_en}) then follows; we used it in the form
(\ref{energybalance:2d}) in connection with our hypothesis of
proportionality of the annihilation and creation rates  to the
dissipation rate. 

\section{Discussion}\label{sec:6}

Now that we have postulated that the rates of STZ creations and 
annihilations are proportional to the rate of energy dissipation and shown 
how to derive  dynamical equations, we will proceed with discussing 
physical mechanisms that can underlie  this hypothesis, and possible
directions to further develop the STZ theory.

The real microscopic picture of plastic deformation is far more 
complicated than what we describe in our model, where the properties 
of material are determined by the behavior of STZs only. 
At present, we can only tell if an STZ exists by observing localized 
atomic rearrangements -- transitions from one STZ state to the other.
But in principle an STZ is a spot where transition is potentially possible.
However, since we judge about presence of an STZ only after the fact 
of transition,
it is impossible to say whether an STZ was annihilated or created if
we did not see its transition; or, even if we saw its transition, it is
impossible to say at what point in time the STZ was created or annihilated.

Thus, for example, it is much easier to understand the energetic
properties related to a transition that already happened. When atoms 
in an STZ rearrange, an additional stress field is created around the 
place of rearrangement. It is in this field that
the plastic energy $\psi$ is stored, and this energy
is in principle recoverable during a reverse transition. 

However, how do we understand the energetic processes related
to the elusive events of STZ creations and annihilations? 
Annihilations are easy to imagine as impossibility of reverse
transition after the initial transition or a series of transitions. 
In this case we can say that the STZ has annihilated and the energy
stored in it has dissipated. Hence we can see a direct connection
between the dissipated energy and annihilation. 

Let us look further. Any transition at low temperatures is a transition 
from a higher
energy state to a lower energy state.
This transition and creation of the stress field around the STZ
is accompanied by dissipation of energy equal to the difference
between the energy levels. This difference, before being absorbed by
thermostat, can cause significant local increase of kinetic energy
and additional atomic rearrangements which, along with
transitions of other STZs, can lead to creations of new STZs and 
annihilations of existing ones. Thus, the energy dissipation is again
related to creations and annihilations. 

Another important problem is to consider other essential degrees 
of freedom describing the structure of material.
As we mentioned earlier, $n_\infty$ can be especially sensitive
to them. In a  theory for elevated temperatures, it is  the 
increasing temperature dependence of $n_\infty$ that gives calorimetric 
characteristics of glass transition. 
An interesting way to introduce a variable describing  disorder in the  
structure of material was offered  in \cite{LANGER:2004}, 
where $n_\infty$ was assumed to depend on that new variable 
instead of directly on the temperature.

In the complex and not yet fully understood picture of the 
microscopic mechanisms underlying plastic deformation in amorphous
solids, the conjecture of proportionality offered in this paper
is the simplest of what can be suggested for STZ creation and
annihilation rates, and it can be useful beyond the current
framework of the low temperature theory.
   
\begin{acknowledgments}
This research was primarily supported by U.S. Department of Energy
Grant No. DE-FG03-99ER45762.  I particularly wish to thank Jim Langer
for constant attention  to this work and  useful comments during
preparation of this manuscript,  and  Lance Eastgate for many useful
suggestions.  I would also like to thank Craig Maloney, Anthony Foglia
and Anael Lemaitre for helpful discussions.
\end{acknowledgments}

\bibliography{allrefs}

\end{document}